\documentclass[preprint,twocolumn,superscriptaddress,]{revtex4}
\usepackage{amssymb,amsmath}
\usepackage{graphicx}
\usepackage{dcolumn}
\usepackage{multirow}
\usepackage{color}
\usepackage{hyperref}
\usepackage{epsfig}
\usepackage{bm}

\newcommand{\unit}[1]{\;\mathrm{#1}}

\begin{document}

\title{\textbf{\fontfamily{phv}\selectfont Thermionic charge transport in CMOS nano-transistors}}
\author{A.~C.~Betz}
\email{ab2106@cam.ac.uk}
\affiliation{Hitachi Cambridge Laboratory, J. J. Thomson Avenue, Cambridge CB3 0HE, United Kingdom}
\author{S. Barraud}
\affiliation{SPSMS, UMR-E CEA / UJF-Grenoble 1, INAC, 17 rue des Martyrs, 38054 Grenoble, France}
\author{Q. Wilmart}
\affiliation{Laboratoire Pierre Aigrain, ENS-CNRS UMR 8551, Universit\'{e}s P. et M. Curie and Paris-Diderot, 24, rue Lhomond, 75231 Paris Cedex 05, France}
\author{B. Pla\c{c}ais}
\affiliation{Laboratoire Pierre Aigrain, ENS-CNRS UMR 8551, Universit\'{e}s P. et M. Curie and Paris-Diderot, 24, rue Lhomond, 75231 Paris Cedex 05, France}
\author{X. Jehl}
\affiliation{SPSMS, UMR-E CEA / UJF-Grenoble 1, INAC, 17 rue des Martyrs, 38054 Grenoble, France}
\author{M. Sanquer}
\affiliation{SPSMS, UMR-E CEA / UJF-Grenoble 1, INAC, 17 rue des Martyrs, 38054 Grenoble, France}
\author{M.~F.~Gonzalez~-~Zalba}
\affiliation{Hitachi Cambridge Laboratory, J. J. Thomson Avenue, Cambridge CB3 0HE, United Kingdom}

\date{\today}

\begin{abstract}
We report on DC and microwave electrical transport measurements in silicon-on-insulator CMOS nano-transistors at low and room temperature. At low source-drain voltage, the DC current and RF response show signs of conductance quantization. We attribute this to Coulomb blockade resulting from barriers formed at the spacer-gate interfaces. 
We show that at high bias transport occurs thermionically over the highest barrier: Transconductance traces obtained from microwave scattering-parameter measurements at liquid helium and room temperature is accurately fitted by a thermionic model. From the fits we deduce the ratio of gate capacitance and quantum capacitance, as well as the electron temperature.  
\end{abstract}
\maketitle
Recent CMOS technology allows for the fabrication of silicon-on-insulator structures of nano-metric size. Depending on fabrication strategy, charge transport in these devices can be quasi-zero- or one-dimensional: Field-effect devices based on narrow silicon channels with spacer regions surrounding the gate can exhibit single-electron-transistor (SET) characteristics \cite{HofheinzAPL2006,RocheAPL2012,ShinNanoLett2011}, whereas gate-all-arround nanowires formed in Si fins show 1D behavior, e.g. conductance quantization due to subband formation \cite{ColingeIEEE2006,YiNanoLetters2011,RazaviehNanoLett2013}.
Silicon nanowires on silicon-on-insulator (SOI) substrates have recently been shown to be promising candidates for future low-power and radio frequency (RF) applications \cite{BarraudIEEE2012,GrenouilletIEEE2012,RazaviehNanoLett2013,RazaviehIEEE2013}. The linearity of the RF response is improved due to the one-dimensional, ballistic charge transport in such devices and the possibility to operate them in the so called quantum capacitance limit, where the gate voltage controls mainly the bands or levels in the transistor channel.
Whether these requirements for RF linearity can also be met in devices that predominantly exhibit SET-like behaviour has not been studied extensively yet.\newline
\begin{figure}[h]
  \includegraphics[width=0.98\columnwidth]{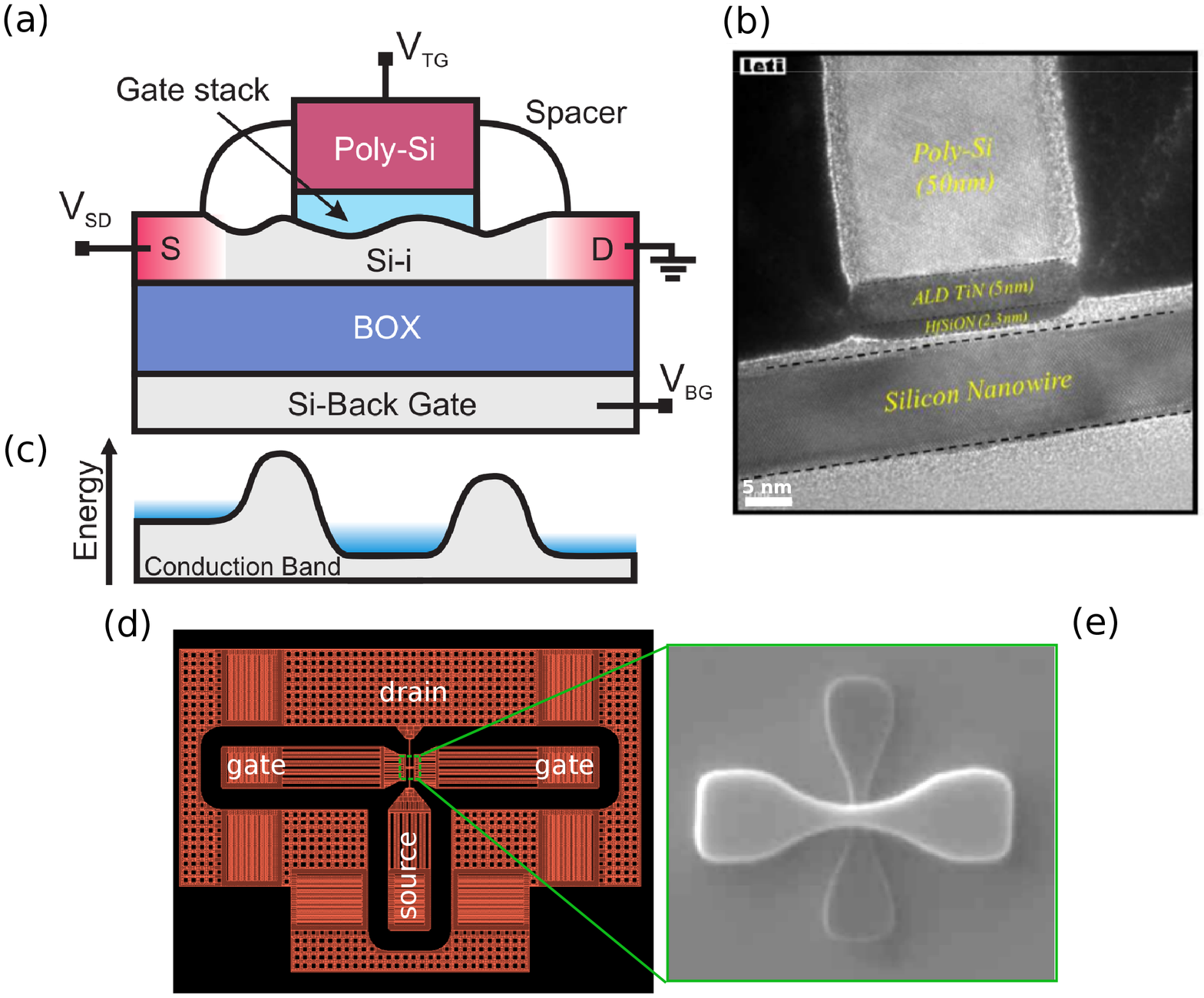}
  \caption{(Color online) Device design and potential profile. (a) Sketch of device cross-section. From bottom to top: Undoped Si handle wafer (green), $145$ nm buried oxide (blue), undoped Si channel of thickness $11\unit{nm}$ (green, width $W=10$ - $60\unit{nm}$) and thin dielectric stack (cyan, $0.8$ nm $\mathrm{SiO_2}$, $2.3$ nm HfSiON) separating top-gate from channel (red, gate lengths $L=44\unit{nm}$, $64\unit{nm}$ and $114\unit{nm}$). Highly doped source and drain separated from gate by SiN spacers (grey).  (b) Transmission electron micrograph of device cross-section as described in (a). (c) Sketch of energy profile along the channel. Potential barriers form at spacer-gate interface. (d) High frequency chip design: Devices are embedded in $50\unit{\Omega}$ adapted coplanar wave-guide with surrounding ground plane. (e) Scanning electron micrograph of highlighted region in (d) during fabrication.}\label{Fig1}
\end{figure}
This letter presents a study of the nature of electron transport in narrow channel CMOS silicon field-effect devices at low and high bias. They consist of a narrow Si channel in silicon-on-insulator substrate with a local top-gate and global back-gate. Spacer elements separate the source and drain electrodes from the top-gate, which in conjunction with surface roughness and remote charges in the gate stack induces two barriers in the potential landscape \cite{VoisinPreparation2013}. We show that at low temperature and low bias electron transport is governed by Coulomb blockade due to these barriers, observable directly e.g. in the source-drain current as a function of bias and top-gate voltage at $\unit{mK}$ temperature.
From microwave scattering-parameter measurements at low temperature we determine a different impact of the barriers depending on source-drain voltage: At low bias both barriers contribute, whereas at high bias transport occurs quasi-thermionically and only the highest of both barriers plays a role. Using a thermionic transport model we are able to reproduce the device's transconductance and deduce the ratio of total capacitance and quantum capacitance as well as the electron temperature. Further microwave measurements show that the devices operate in the hot electron regime at high bias up to room temperature.
\newline
Devices were fabricated in fully depleted SOI substrate at LETI facilities. First the active regions were patterned by etching the SOI layer above the $145\unit{nm}$ buried oxide (BOX) forming undoped Si channels of thickness $t=12\unit{nm}$. After short oxidation of the channel ($0.8\unit{nm}$) the gate stack is formed ($1.9\unit{nm}$ HfSiON, $5\unit{nm}$ TiN and $50\unit{nm}$ poly-Si) and etched (see Fig.\ref{Fig1}(e)). Silicon nitride spacers of length $11\unit{nm}$ were deposited on both sides of the gate and the source-drain contacts raised by epitaxial growth of Si ($18\unit{nm}$). Both source and drain were then highly doped by extension implantation and activation annealing. Fig.\ref{Fig1}(a) and (b) show a sketch and a TEM micrograph of a typical device. Similar to \cite{RocheAPL2012}, the devices have a non-overlap profile due to the $\mathrm{Si_3N_4}$ spacers. The doping gradient under the spacers together with surface roughness and remote charges in the gate stack induce potential barriers as sketched in Fig.\ref{Fig1}(c). In a last step the devices were silicided (NiPtSi). All devices presented in this letter are embedded in a $50\unit{\Omega}$ adapted coplanar wave guide for RF measurements (see Fig.\ref{Fig1}(d)). Gate lengths $L_g$ and channel widths $W$ of devices present in this letter are shown in Table I.\newline
\begin{table}\centering
\begin{tabular*}{0.95\columnwidth}{@{\extracolsep{\fill} } l|cc||l|cc}\toprule
\emph{device} & \emph{$L_g\unit{(nm)}$} & \emph{$W\unit{(nm)}$} & \emph{device} & \emph{$L_g\unit{(nm)}$} & \emph{$W\unit{(nm)}$} \\ \hline
\emph{RFM2-1} & 44 & 10 & \emph{RFM4-3a} & 64 & 30 \\
\emph{RFM1-2} & 44 & 30 & \emph{RFM4-3b} & 64 & 30\\ 
\emph{RFM1-4} & 64 & 10 & \emph{RFM3-1} & 114 & 10\\ 
\hline
\hline
\end{tabular*}\label{TableDevices}
\caption{Device dimensions.}
\end{table}

\begin{figure}[h]
  \includegraphics[width=0.98\columnwidth]{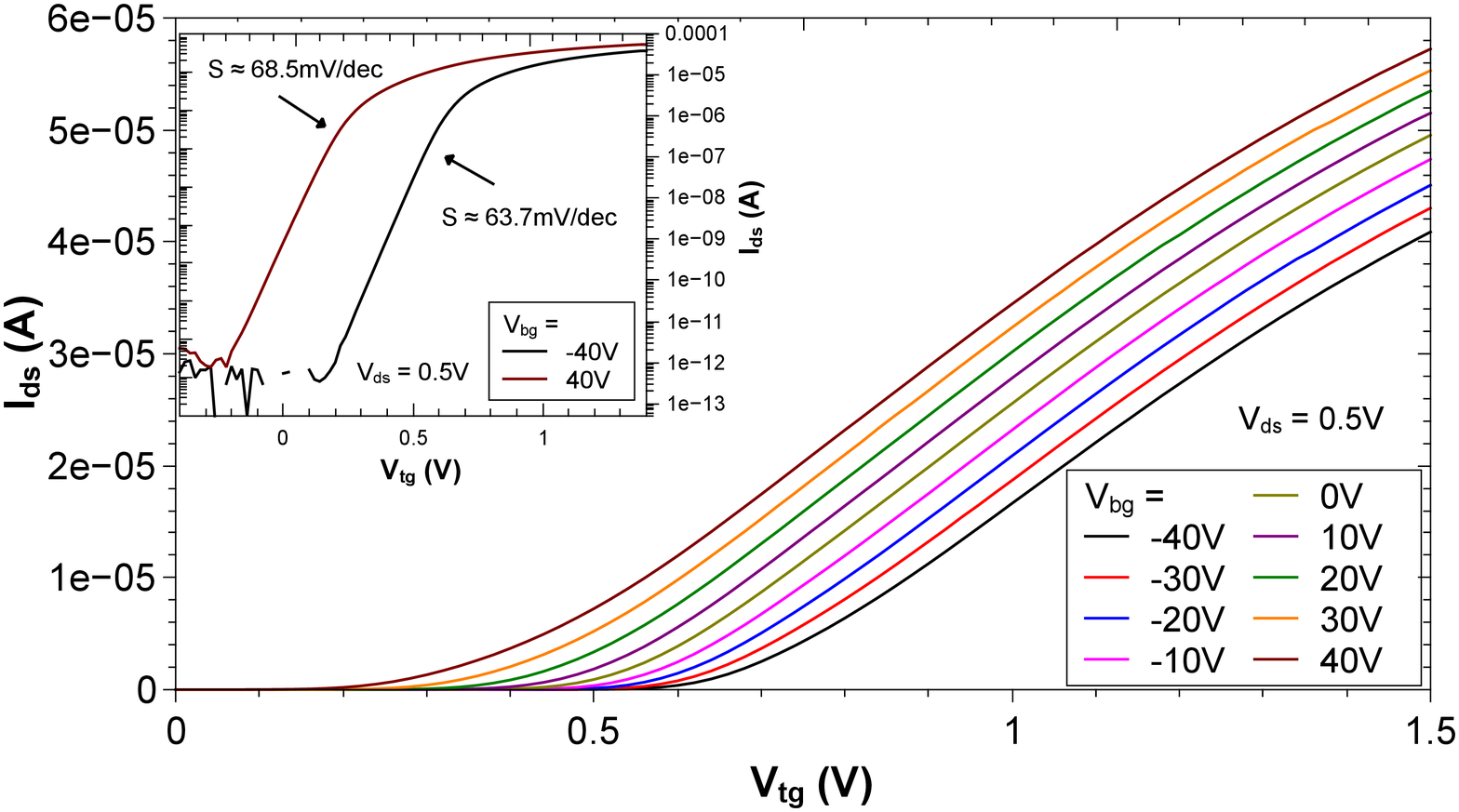}
  \caption{(Color online) Device RFM1-2 at $T_0=300\unit{K}$. Main panel: Source-drain current $I_{ds}$ as function of top-gate voltage $V_{tg}$ at high source-drain bias for back-gate voltages $V_{bg}=-40\unit{V}$ - $40\unit{V}$. Inset: $log_{10}(I_{ds})$ as function of $V_{tg}$ at $V_{bg}=\pm 40\unit{V}$. The subthreshold swing S is calculated from the inverse of $d (log_{10}I_{ds})/d V_{tg}$.}\label{Fig2}
\end{figure}
DC and RF probe station measurements were carried out at room temperature and liquid helium temperature in a \textit{Janis} variable temperature probe station. 
The RF response was probed with a vector network analyzer at frequencies $f\leq 20$GHz: After a short-open-load-through calibration, we obtained the scattering parameters $S_{ij}(\omega)\, ,\, i,j=1,2$ for each pair of gate voltages and converted them to admittance parameters $Y_{i,j}(\omega)$ \cite{PallecchiAPL2011}. The conversion is favourable since parasitic, parallel elements can then simply be subtracted \cite{Pozar}. We use the device's off-state as dummy signal and obtain the de-embedded admittance $Y_{DUT}(\omega,V_{bg},V_{tg}) = Y_{exp.}(\omega,V_{bg},V_{tg}) - Y_{off}(\omega,V_{bg},V_{tg}\ll V_{th})$, where $V_{th}$ is the threshold voltage, $V_{tg}$ the top-gate and $V_{bg}$ the back-gate voltage. Note that in the following the subscript \textit{DUT} is omitted and $Y_{ij}$ refers to the de-embedded admittance signal. DC traces are recorded simultaneously with the RF measurements.\newline
In Fig.\ref{Fig2} we present first of all the DC current $I_{ds}$ of device RFM1-2 ($W=30\unit{nm}$, $L_g=44\unit{nm}$) at $T_0=300\unit{K}$ as a function of top-gate voltage for various back-gate voltages. The device shows a good transistor behavior with off-state current in the $\mathrm{pA}$ region for $V_{ds}=0.5\unit{V}$ and a subtreshold slope $S\simeq 64$ - $69\unit{mV/dec}$ (see Fig.\ref{Fig2}), close to the theoretical limit of $k_B T ln(10) /q \simeq 60\unit{mV/dec}$ for thermally activated transport at $T=300\unit{K}$ \cite{Sze}. Here, $k_B$ is the Boltzmann constant and $q$ the electron charge. As previously reported by \cite{RocheAPL2012}, the threshold voltage $V_{tg}$ shifts with back-gate voltage. 
\begin{figure}[h]
  \includegraphics[width=0.98\columnwidth]{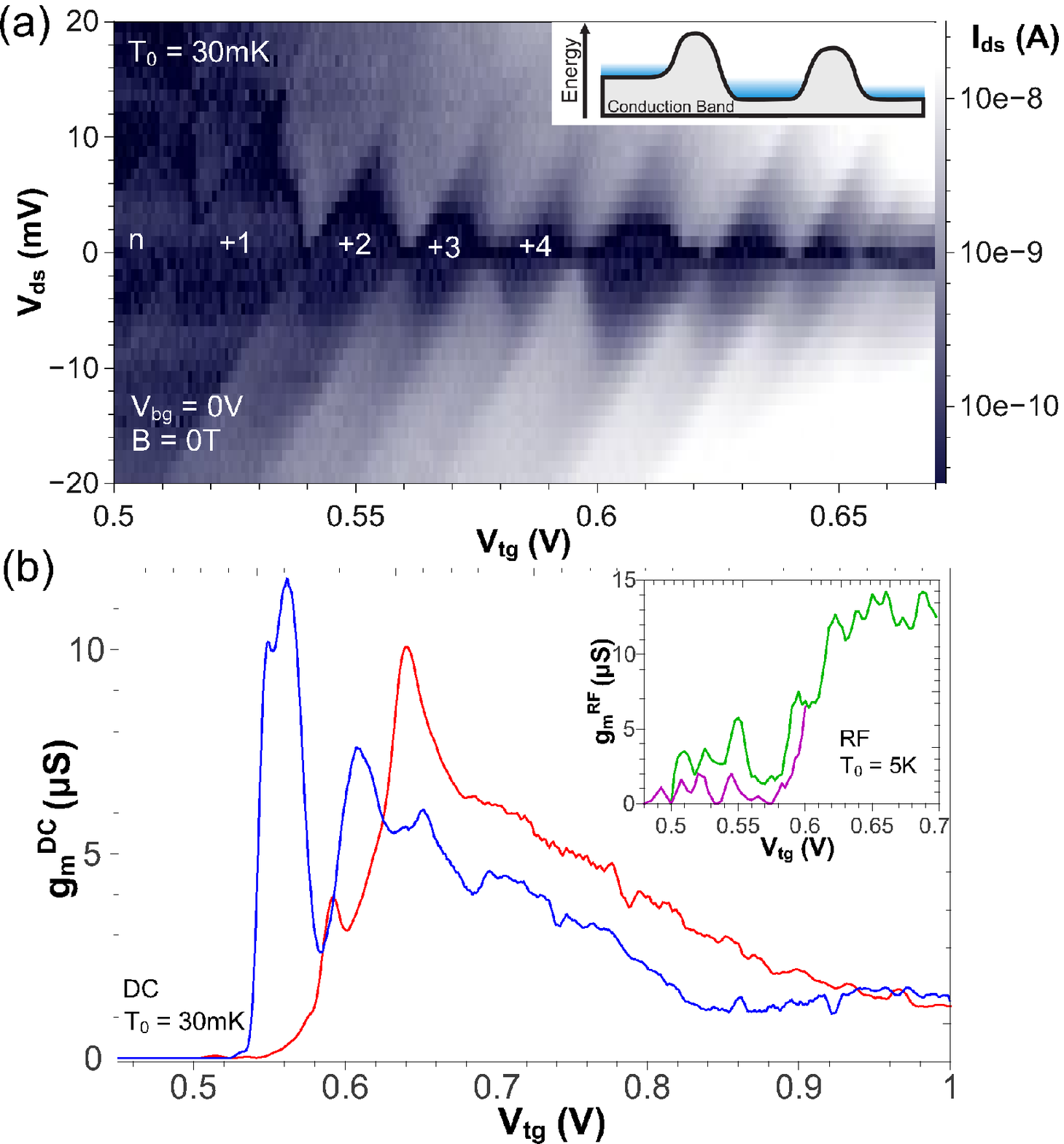}
  \caption{(Color online) (a) Source-drain current $I_{ds}$ as function of $V_{ds}$ and $V_{tg}$ in device RFM1-4 ($W=10\unit{nm}$, $L_g=64\unit{nm}$) at $T_0=30\unit{mK}$ and $V_{bg}=0\unit{V}$. Coulomb diamonds are clearly visible in the $V_{tg}$ range just below turn-on. Inset: Sketch of the potential landscape due to the barriers at the spacer-gate interfaces at low bias.
(b) main panel: DC transconductance  $g_m^{DC}\,(V_{tg})=\partial I_{ds}/\partial V_{tg}$ in devices RFM4-3a (blue)  and RFM4-3b (red) at $T_0=30\unit{mK}$ and $V_{ds}=25\unit{mV}$. Inset: RF transconductance $g_m^{RF}=\Re(Y_{21})$ of devices RFM3-1 (green) and RFM2-1 (purple) at $T_0=5\unit{K}$ and $V_{ds}=5\unit{mV}$ (RFM3-1) and $10\unit{mV}$ (RFM2-1), respectively.} \label{Fig3}
\end{figure}
We turn now to characterizing the nature of charge transport in our devices. As described in the introduction of this letter, charge transport through a narrow silicon channel can in general occur via one-dimensional channels \cite{RazaviehNanoLett2013} or through the levels of a single-electron-transistor-like structure \cite{RocheAPL2012}. In order to discriminate between the two mechanisms, we have carried out DC measurements in a dilution fridge setup at $T_0=30\unit{mK}$ and RF probe station measurements at $T_0=5\unit{K}$. Fig.\ref{Fig3}(a) is a colorscale plot of the current $I_{ds}$ in sample RFM1-4 just below the turn-on voltage of the transistor at $T_0=30\unit{mK}$. Clear Coulomb diamonds are visible, indicating the presence of two barriers and subsequent conductance quantization as depicted in Fig.\ref{Fig1}(c) and the inset to Fig.\ref{Fig3}(a).
Fig.\ref{Fig3}(b) shows the transconductance $g_m$ of devices RFM4-3, RFM3-1 and RFM2-1 at low bias and $T_0=5\unit{K}$ and $30\unit{mK}$, respectively. The DC transconductance is calculated from the DC current as $g_m^{DC}= \partial I_{ds}/\partial V_{tg}$. At RF it can directly be read from the forward admittance as $g_m^{RF}=\Re(Y_{21})$. Similar to the device of Fig.\ref{Fig3}(a), RFM4-3a and RFM4-3b show Coulomb blockade at low $V_{ds}$ with a charging energy of about $15\unit{meV}$ (not shown here). The DC transconductance reflects this, as can be seen from the oscillations of $g_m$ at $V_{ds}=25\unit{mV}$ in Fig.\ref{Fig3}(b). In addition, the DC transconductance remains flat below threshold. Above threshold we observe two $g_m^{DC}$ peaks and a steady decline thereafter. The RF counterparts of RFM3-1 and RFM2-1 were obtained at $T_0=5\unit{K}$ and $V_{ds}=5\unit{mV}$ and $10\unit{mV}$, respectively. Consequently these traces exhibit several peaks below threshold as function of $V_{tg}$ with a gate spacing of $\Delta V_{tg}\simeq 15$ - $22\unit{mV}$ for RFM2-1 (purple). For the device of longest gate length, RFM3-1 (green), peaks below threshold are less pronounced, but prevail beyond threshold, similar to $g_m^{DC}$ of devices RFM4-3a and b. We attribute the additional small variations, e.g. at $V_{tg}\simeq 0.6\unit{V}$, to increased disorder in the device. Overall, we infer that low bias transport is governed by barriers formed at the spacer-gate interfaces. Their effect manifests in both DC and RF measurements. 
\newline
The picture changes when the source-drain bias is raised to higher values: Fig.\ref{Fig4}(a) shows the transconductance extracted from low temperature RF scattering-parameter measurements in device RFM2-1. Contrary to the low bias case of Fig.\ref{Fig3}(b), $g_m$ now shows only one broad peak as function of $V_{tg}$. Owing to the high $V_{ds}$, i.e. large potential drop between source and drain, transport is now dominated by only one barrier and we can fit data by a simple 1D nanotransistor model, as outlined in \cite{ChasteAPL2010}. Here, the transistor is described as a 1D channel with one classical barrier between the leads. A corresponding potential landscape is shown as inset of Fig.\ref{Fig4}(a). The model takes into account a potential as highlighted in red. Only electrons with energies high enough to clear the barrier contribute to the charge transport in this model. Transport over the barrier is thus thermionic and one can derive the corresponding transconductance \cite{ChasteAPL2010} 
\begin{equation}\label{gmThermionic}
g_m=\alpha \beta \cdot\frac{2e^2}{h}[f_s(\Phi)-f_d(\Phi)],
\end{equation}
where $\beta= C_g/C_q$. $C_g=(C_{ox}^{-1}+C_q^{-1})^{-1}$ and $C_q$ are the total gate and the quantum capacitance \cite{JohnJAP2004}, respectively; $\alpha$ is an additional factor accounting for residual diffusive charge transport over the barrier. The contacts are modelled by the Fermi distributions $f_s(\epsilon,T_e)$ and $f_d(\epsilon,T_e) = f_s(\epsilon+eV_{ds},T_e)$, where $T_e$ is the electron temperature. All three parameters, $\alpha$, $\beta$ and $T_e$, depend on source-drain voltage and back-gate bias. Assuming that the barrier height is controlled by $V_{tg}$, i.e. $\Phi(V_{tg}) = const. - qV_{tg}\beta$, as well as $\epsilon \ll \Phi$ \cite{ChasteAPL2010} and using $\alpha$, $\beta$, $T_e$ as fitting parameters, we obtain the dashed lines shown in Fig.\ref{Fig4}(a) and (b). The good agreement between data and fit - performed for each $V_{bg}$ - shows that transport is dominated by only the highest barrier in the high bias regime and can be regarded as quasi one-dimensional. 
At low temperature the transconductance peak is more pronounced than at room temperature and we find from the fits that the device works in the hot electron regime: The electron temperature $T_e$, shown in Fig.\ref{Fig4}(c) as function of $V_{bg}$, is much higher than the substrate temperature $T_0=5\unit{K}$. Fig\ref{Fig4}(c) also presents the extracted ratio $\beta$ of total gate to quantum capacitance. On average $\beta\simeq 0.6$, i.e. $C_q\simeq C_{ox}/2$. The device hence operates in between the classical, charge controlled limit ($\beta\rightarrow 0$) and the quantum capacitance limit ($\beta\rightarrow 1$) \cite{RazaviehIEEE2013}. Similar to the RF transconductance, one can obtain the total gate capacitance from RF Y-parameter measurements as $Y_{11}(\omega))= j\omega(C_{gs}+C_{gd})$. We find $C_{tot}\simeq 600$ - $50\unit{aF}$ for $V_{bg}=0$ - $80\unit{V}$. Residual diffusive transport remains moderate with an average "ballisticity" factor $\alpha\approx 0.77$ (not shown here). At room temperature, we obtain a noisier RF response and thus larger uncertainty in the fitting procedure. Data can still be modelled by Eq.\ref{gmThermionic}, i.e. transport is still thermionic. On average, the electron temperature remains at a similar level as in the low temperature measurements (see Fig.\ref{Fig4}(d)), as does the capacitance ratio $\beta$. Residual diffusive transport however increases at room temperature and we now obtain on average $\alpha\approx 0.6$ (not shown here).
\newline
\begin{figure}[h]
  \includegraphics[width=0.98\columnwidth]{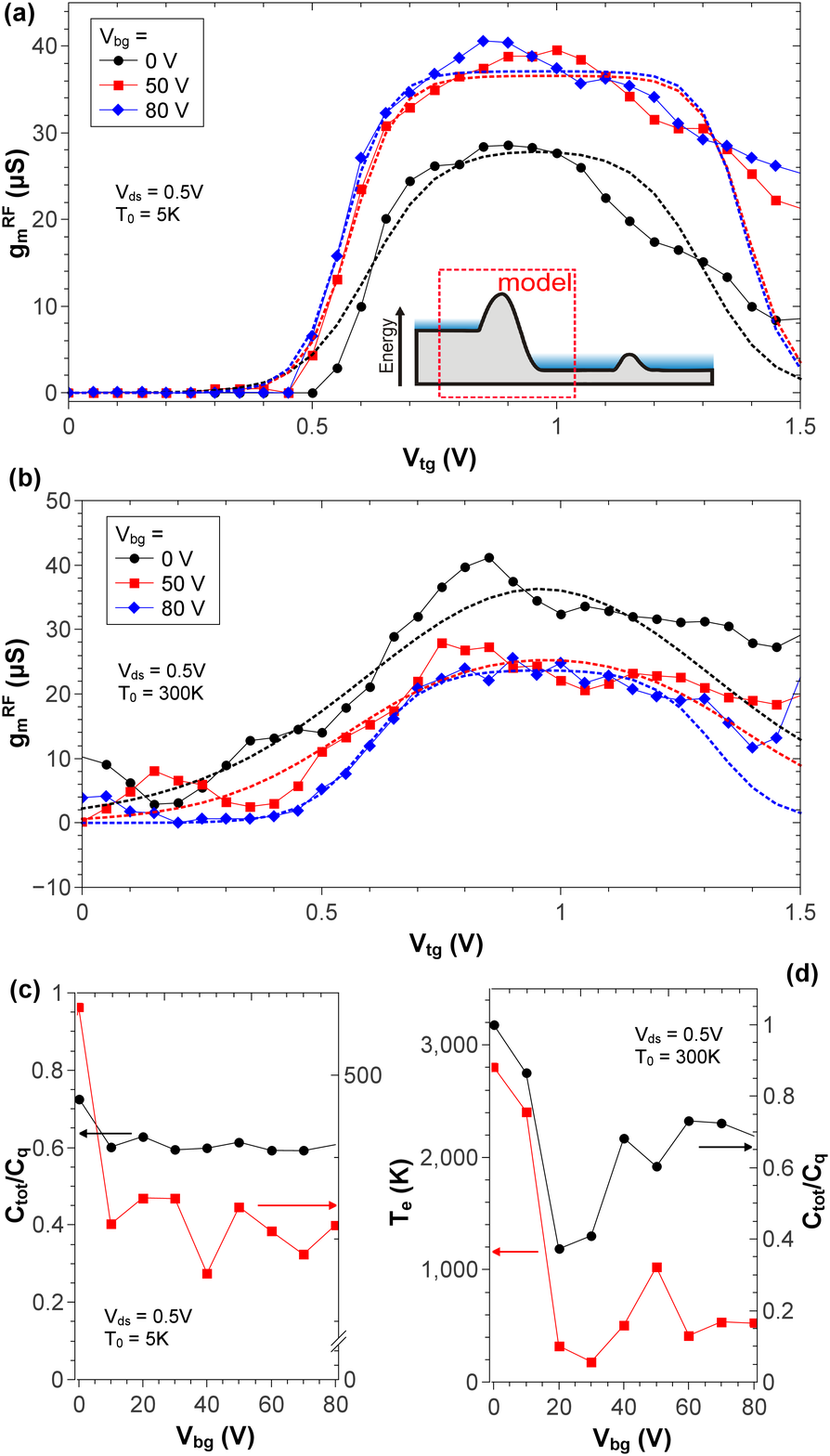}
  \caption{(Color online) Device RFM2-1 
  (a) RF transconductance at low temperature and high bias. Symbols show the measured values at three different back-gate voltages. Dashed lines are fits to the data following the thermionic model of Eq.\ref{gmThermionic}. The high bias potential landscape is sketched as inset. (b) RF transconductance at room temperature and high bias. As in the $T=5\unit{K}$ case data can be accurately fitted by the thermionic model. (c) and (d) Electron temperature $T_e$ and capacitance ratio $\beta$ at $T_0=5\unit{K}$ and $300\unit{K}$ obtained from the thermionic fits in (a) and (b).}\label{Fig4}
\end{figure}
In conclusion, we have studied the nature of charge transport at low and high bias in narrow channel SOI field-effect devices. We show that at low bias transport is governed by barriers emerging in the undoped region under the spacer elements, whereas at high bias transport occurs quasi-one-dimensional over the highest barrier. In the latter case the device's behaviour can be described with a thermionic transport theory and we accurately model the device transconductance. The devices operate in the thermionic transport regime from liquid helium to room temperature.

The authors thank D.A. Williams and A.J. Ferguson for fruitful discussion. The research leading to these results has been supported by the European Community’s seventh Framework under the Grant Agreement No. 318397. The samples presented in this work were designed and fabricated by the TOLOP project partners, http://www.tolop.eu.


\end{document}